\documentclass[10pt]{amsart}
\usepackage{amsmath}
\usepackage{latexsym}
\usepackage{amsfonts}
\usepackage{amssymb}
\usepackage{amsthm}
\usepackage{bbm,dsfont}
\usepackage{graphicx}

\newtheorem{proposition}{Proposition}

\theoremstyle{definition}
\newtheorem{example}{Example}

\newcommand{\R}{\mathbb R} %real
\newcommand{\half}{\tfrac{1}{2}} %half

\newcommand{\ip}[2]{\left\langle\,#1\,|\,#2\,\right\rangle} %inner product
\newcommand{\no}[1]{\left\|#1\right\|} %norm
\newcommand{\tr}{\textrm{tr}} %trace
\newcommand{\id}{\mathbbm{1}} %identity operator
\newcommand{\nul}{\mathbbm{O}} %null operator
\newcommand{\fii}{\varphi}

\newcommand{\va}{\mathbf{a}} %a
\newcommand{\vb}{\mathbf{b}} %b
\newcommand{\vg}{\mathbf{g}} %g
\newcommand{\vu}{\mathbf{u}} %u
\newcommand{\vr}{\mathbf{r}} %r
\newcommand{\vn}{\mathbf{n}} %n
\newcommand{\vm}{\mathbf{m}} %m
\newcommand{\vx}{\mathbf{x}} %x
\newcommand{\vsigma}{\boldsymbol{\sigma}} %sigma
\newcommand{\vnull}{\mathbf{0}}%null vector

\newcommand{\Aaa}{A(\alpha,\va)} %(\alpha,a)
\newcommand{\A}{\mathcal{A}}%generic observable
\newcommand{\B}{\mathcal{B}}%generic observable
\newcommand{\E}{\mathcal{E}}%generic observable
\newcommand{\F}{\mathcal{F}}%generic observable
\newcommand{\G}{\mathcal{G}}%generic joint observable
\newcommand{\Ea}{\mathcal{E}^{1,\mathbf{a}}} %(1,a)
\newcommand{\Eb}{\mathcal{E}^{1,\mathbf{b}}} %(1,b)
\newcommand{\Ead}{\mathcal{E}^{1,\mathbf{a}'}} %(1,a')
\newcommand{\Ebd}{\mathcal{E}^{1,\mathbf{b}'}} %(1,b')
\newcommand{\Eaa}{\mathcal{E}^{\alpha,\mathbf{a}}} %(\alpha,a)
\newcommand{\Ebb}{\mathcal{E}^{\beta,\mathbf{b}}} %(\beta,b)
\newcommand{\Esa}{\mathcal{E}^{1,\hat\va}} %(1, a)
\newcommand{\En}{\mathcal{E}^{1,\mathbf{n}}} %(1,n)
\newcommand{\Em}{\mathcal{E}^{1,\mathbf{m}}} %(1,m)
\newcommand{\Emn}{\mathcal{E}^{1,-\mathbf{m}}} %(1,-m)
\newcommand{\Ha}{\mathcal{H}}

\newcommand{\D}{\mathfrak{D}}%distance and error
\newcommand{\U}{\mathfrak{U}}%unsharpness

\newcommand{\sa}{\sigma_A}%spectrum of A
\newcommand{\sap}{\sigma_{A'}}%spectrum of A'
\newcommand{\saap}{\sigma_{AA'}}%spectrum of AA'
\renewcommand{\S}{\mathfrak{S}}%sharpness of ...
\newcommand{\W}{\mathfrak{W}}%width

\newcommand{\bB}{\mathbf{B}}
\newcommand{\cC}{\mathcal{C}}
\newcommand{\uc}{\lor}
\newcommand{\dc}{\land}

\begin{document}

\title{Approximate Joint Measurements of Qubit Observables}

\author{Paul Busch}
\address{Department of Mathematics, University of York, York, UK}
\email{pb516@york.ac.uk}

\author{Teiko Heinosaari}
\address{Research Center for Quantum Information, Slovak Academy of Sciences, Bratislava, Slovakia, and Department of Physics, University of Turku, Finland}
\email{heinosaari@gmail.com}

\begin{abstract}
Joint measurements of qubit observables have recently been studied in conjunction with quantum information processing tasks such as cloning. Considerations of such joint measurements have until now been restricted to a certain class of observables that can be characterized by a form of covariance. Here we investigate conditions for the joint measurability of arbitrary pairs of qubit observables. For pairs of noncommuting sharp qubit observables, a notion of approximate joint measurement is introduced. Optimal approximate joint measurements are shown to lie in the class of covariant joint measurements. The marginal observables found to be optimal approximators are generally not among the coarse-grainings of the observables to be approximated. This yields scope for the improvement of existing joint measurement schemes. Both the quality of the approximations and the intrinsic unsharpness of the approximators are shown to be subject to Heisenberg-type uncertainty relations.
\end{abstract}

\date{}
\maketitle

%\tableofcontents

\section{Introduction}\label{sec:introduction}

In recent years there has been an increasing interest in the question of joint measurability of noncommuting quantum observables, both from a foundational
\cite{OQP,FQMEA,AnBaAs05,SoAnBaKi05,Kar-etal06,JaBo06}
and quantum information theoretical
\cite{DArSa00,DArMaSa01,BrAnBa06,FePa07} perspective. The connection of this issue with certain impossible tasks in quantum mechanics, such as universal copier and Bell's telephone, is lucidly explained in \cite{Werner01}. Since two observables represented as selfadjoint operators do not have a joint observable if they do not commute, it is necessary in such cases to understand joint measurability in a wider sense. 

As intuitively understood by Heisenberg already in 1927 \cite{Heisenberg27}, one has to allow for a degree of imprecision in order to make room for a notion of joint measurement of noncommuting observables. This idea can be appropriately investigated if the wider class of observables represented as positive operator measures (POMs) is taken into consideration. Projection valued measures among the POMs correspond to the standard observables represented as selfadjoint operators; they are called {\em sharp observables}.

In the class of POMs, there are pairs of noncommuting observables that possess a joint observable, which thus has these two observables as its marginals. Commutativity is necessary for joint measurability if at least one of the POMs is a sharp observable, but generally commutativity is not required. A joint measurement of two POMs $\E^1$ and $\E^2$ can be regarded as an {\em approximate joint measurement} of two noncommuting observables $\A$ and $\B$ if $\E^1,\E^2$ are close (in some suitable sense) to $\A,\B$, respectively.

The problem of approximate joint measurements of position and momentum has been treated comprehensively in related publications \cite{Werner04b,BuHeLa06,BuPe06}. The case of observables with discrete spectra requires somewhat different concepts and will be treated in the present paper for the case of joint measurements of qubit observables. We will introduce an appropriate measure of the quality of the approximation of one observable by another observable. It will then be shown that the quality of approximations in an approximate joint measurement of two sharp observables is limited if these observables do not commute. This limitation can indeed be formulated rigorously as a form of Heisenberg uncertainty relation.

One factor limiting the accuracy in an approximate joint measurement of noncommuting sharp observables is the fact that the approximating marginal observables must have a sufficient degree of {\em intrinsic unsharpness} as a consequence of their joint measurability. Hence there is yet another form of Heisenberg uncertainty relation for appropriately defined degrees of unsharpness in joint measurements. The distinction between the relational feature of inaccuracy (distance between two POMs) and the intrinsic property of unsharpness (of an individual POM) was until now blurred due to the fact that joint measurements were considered in which the marginals were coarse-grained versions of the sharp observables to be approximated; in such cases the intrinsic unsharpness and the inaccuracy are interconnected.

A theory and first models of approximate joint measurements of qubit observables
were presented for special cases in \cite{Busch86,Busch87}. In those works and all subsequent developments, only a restricted class of joint measurements was used to approximate two sharp spin components. Here this restriction will be lifted, thereby allowing one to determine optimal joint measurements and to formulate uncertainty relations that can be used to characterize the optimal cases.

The paper is organized as follows. In Section \ref{sec:simple} we
review the condition of joint measurability  of two simple
observables  (i.e. observables representing yes-no measurements),
and give a precise definition of the approximate joint measurability
of two observables. These conditions and concepts are investigated
in Sections~\ref{sec:qubit}-\ref{sec:approx-joint} in the case of
qubit observables. Section~\ref{sec:conclusion} gives our
conclusions and an outlook.

\section{Simple observables and their (approximate) joint measurability}\label{sec:simple}

\subsection{Effects and observables}\label{sec:effects}

The general definition of an observable $\A$ as a positive operator
measure (POM) reduces, in the case of measurements with finitely
many outcomes $\omega_i$, to the specification of a map
$\omega_i\mapsto\A_i$, where the $\A_i$ are {\em effects}, that is,
positive operators satisfying the ordering
relation\footnote{Relation $A\leq B$ for two selfadjoint operators
$A$ and $B$ means that $\ip{\psi}{A\psi}\leq\ip{\psi}{B\psi}$ for
every vector $\psi$.} $\nul \le \A_i\le\id$. (Here $\nul,\id$ are
the null and unit operators, respectively.) Together with any state
(density operator) $T$, $\A$ determines a probability distribution
over the outcomes of $\A$ via the trace formula, $\omega_i\mapsto
\tr[T\A_i]$. The additivity and normalization of probability
distributions is ensured by the condition $\sum_i\A_i=\id$.

A {\em simple observable} is one that represents a measurement with
two possible outcomes; it is given as a POM with two values and
associated effects,
\begin{equation}
\mathcal{A}:\quad \omega_+\mapsto \A_+,\quad \omega_-\mapsto \A_-.
\end{equation}
Normalization entails that $\A_+ + \A_-=\id$, so that a simple observable
is commutative. We note that any effect $A$ together with its 
complement effect $A':=\id- A$ defines a class of simple
observables, distinguished only by their outcome sets
$\{\omega_+,\omega_-\}$.\footnote{For clarity we denote observables with
script capital letters, while effects are denoted with italic letters.}

As noted in the introduction, an observable (POM) is {\em sharp} if its effects are
all projections. Otherwise an observable is called {\em unsharp}. 
A measure of the intrinsic unsharpness of an effect $A$ and thus  of
the associated simple observable $\A$ that is independent of the
outcomes of $\A$ is obtained as follows. Let $\sa$ denote the
spectrum of an effect $A$, then the {\em spectral width} of $A$ is
the length of the smallest closed interval containing $\sa$, that
is: $\W(\sa):=\max\sa-\min\sa=\no{A}+\no{A'}-1$. The {\em sharpness}
of $A$ is defined as
\begin{equation}\label{eqn:sharpness-def}
\S(A):=\W(\sa)-\W(\saap).
\end{equation}
The operator $AA'$ can be written as $AA'=A^{\half}A'A^{\half}$.
This shows that $AA'$ arises from the sequential L\"uders
measurement of the complement effects $A$ and $A'$, and this is a
motivation for the formula (\ref{eqn:sharpness-def}); see \cite{Busch07} for further discussion.

Since $\W(\sa)=\W(\sap)$, it follows that $\S(A')=\S(A)$. On can prove the following facts about the sharpness\footnote{See \cite{Busch07}. These statements are proved for qubit observables in Subsection \ref{sec:sharpness-distance}.}:
$\S(A)\in[0,1]$; $\S(A)=0$ exactly when $A$ is a
trivial effect (i.e. $A=k\id$ for some $0\leq k\leq 1$);
$\S(A)=1$ exactly when $A$ is a nontrivial projection. These are properties one would expect any measure of sharpness to possess: the measure should single out the perfectly sharp effects and the trivial effects.

The {\em sharpness} and {\em unsharpness} of a simple observable $\A$ may now be defined as
\begin{eqnarray}
\S(\A) &:=& \S(\A_+)=\S(\A_-), \\
\U(\A) &:=& 1-\S(\A)^2. \label{eqn:u}
\end{eqnarray}

\subsection{Joint measurability}\label{sec:joint}

Two observables are jointly measurable if there is a measurement
scheme that allows the determination of the values of both
observables. This means that the POM representing that joint
measurement contains the two observables as marginals. In this way
it is ensured that there is a joint probability distribution for
each state. We spell out this definition\footnote{For the general
definition of joint measurability and a detailed discussion on this
topic, see e.g. \cite{Lahti03} and references given therein.} for
the case of a pair of simple observables.

Two simple observables $\E^1$ and $\E^2$ are \emph{jointly
measurable} if there is an observable $\G:\omega_{ij}\mapsto \G_{ij}$,
$i,j=\pm$, such that
\begin{equation}\label{simple-jm}\begin{split}
\E^1_+=\G_{++}+\G_{+-},&\qquad \E^1_-=\G_{-+}+\G_{--},\\
\E^2_+=\G_{++}+\G_{-+},&\qquad \E^2_-=\G_{+-}+\G_{--}.
\end{split}\end{equation}
In this case the observables $\E^1$ and $\E^2$ are the
\emph{marginals} of $\G$, and we also say that $\G$ is a \emph{joint
observable} for $\E^1$ and $\E^2$. The outcomes $\omega_{ij}$ of
$\G$ could be taken to be (or replaced by) the pairs
$(\omega_i,\omega_j)$.

The joint measurability of two simple observables $\E^1$ and $\E^2$
is equivalent to the statement \cite{SEO} that there exists an
operator $\G_{++}$ satisfying the  following set of operator
inequalities:
\begin{equation}\label{simple-coex}
\begin{split}
\nul\le\G_{++},\quad
\G_{++}&\le\E^1_+,\quad
\G_{++}\le\E^2_+,\\
\E^1_+ + \E^2_+ - \id&\le\G_{++}.
\end{split}
\end{equation}
In fact, these inequalities ensure that the following four operators are effects:
\begin{equation}\label{simple-joint}\begin{split}
&\G_{++},\quad \G_{+-}\equiv \E^1_+ - \G_{++},\quad \G_{-+}\equiv\E^2_+ - \G_{++},\\
&\G_{--}\equiv\id-\G_{++}-\G_{+-} - \G_{-+} = \id-\E^1_+ - \E^2_+ + \G_{++}.
\end{split}\end{equation}
It is straightforward to verify that equations (\ref{simple-jm}) hold and hence, these effects define a joint observable $\G$ for $\E^1$ and $\E^2$.

The joint measurability condition for two simple observables can be interpreted
as the requirement that the intersection of four cones in the set of effects is
nonempty. The order relation $A\le B$ for two selfadjoint operators is equivalent to
either of $B-A\ge\nul$ and $A-B\le\nul$. The condition $A\ge\nul$ defines a convex cone in the real vector space of selfadjoint operators.\footnote{This means that whenever $A,B\ge \nul$, then $tA+(1-t)B\ge\nul$ for any $t\in [0,1]$.} Thus we can define the upward and
downward cones of a selfadjoint operator $A$ as
$\cC^\uc(A):=\{B: A\le B\}$ and $\cC^\dc(A):=\{B:A\ge B\}$. The joint measurability condition for observables $\E^1$ and $\E^2$ now reads:
\begin{equation}\label{eqn:cone-condition}
\cC^\uc(\nul)\cap\cC^\dc(\E^1_+)\cap\cC^\dc(\E^2_+)\cap\cC^\uc(\E^1_++\E^2_+-\id)\ne\emptyset.
\end{equation}

\begin{example}[Trivial cases of joint measurability]\label{ex:trivial-coex}
In the following four cases, joint measurability falls out trivially:\\
(a) $\E^1_+\ge\E^2_+$: put $\G_{++}=\E^2_+$, $\G_{+-}=\E^1_+-\E^2_+$, $\G_{-+}=\nul$, $\G_{--}=\id-\E^1_+$.\\
(b) $\E^1_+\le\E^2_+$: put $\G_{++}=\E^1_+$, $\G_{+-}=\nul$,
$\G_{-+}=\E^2_+-\E^1_+$, $\G_{--}=\id-\E^2_+$.\\
(c) $\E^1_+\ge \E^2_-$: put $\G_{++}=\E^1_++\E^2_+-\id$,  $\G_{+-}=\E^2_-$, $\G_{-+}=\E^1_-$,
$\G_{--}=\nul$.\\
(d) $\E^1_+\le \E^2_-$:  put $\G_{++}=\nul$, $\G_{+-}=\E^1_+$, $\G_{-+}=\E^2_+$, $\G_{--}=\id-\E^1_+-\E^2_+$.
\end{example}

We conclude that interesting (i.e. nontrivial) cases arise when
$\E^1_+-\E^2_+$ and $\E^1_++\E^2_+-\id$
are neither $\le \nul$ nor $\ge \nul$. In terms of a joint observable $\G$, nontrivial cases are exactly those in which $\G_{ij}\neq\nul$ for every $i,j=\pm$.

Another simple instance of joint measurability arises from commutativity.

\begin{example}[Mutually commuting observables]\label{ex:com-coex}
If $\E^1$ and $\E^2$ commute mutually in the sense that $\E^1_i\E^2_j=\E^2_j\E^1_i$ for every $i,j=\pm$, then they are jointly measurable. In this case the formula $\G_{ij}=\E^1_i\E^2_j$ defines a joint observable.
\end{example}

It is instructive to prove the following well-known proposition, which supplements Example \ref{ex:com-coex}. (Its statement is valid also for observables that are not simple, and then the proof requires only minor changes.)

\begin{proposition}\label{prop:com}
Let $\E^1$ and $\E^2$ be simple observables which are jointly measurable. If one of them is a sharp observable, then they commute and the unique joint observable $\G$ is of the product form $\G_{ij}=\E^1_i\E^2_j$.
\end{proposition}

\begin{proof}
Let, for instance, $\E^1$ be a sharp observable and suppose that $\G$ is a joint observable for $\E^1$ and $\E^2$. Since $\G_{ij}\le\E^1_i$, one obtains $\E^1_i\G_{ij}=\G_{ij}\E^1_i=\G_{ij}$. This shows also that $\E^1_i\G_{-ij}=(\id-\E^1_{-i})\G_{-ij}=\nul$ and similarly $\G_{-ij}\E^1_i=\nul$. It follows that
\begin{equation}
\E^1_i\E^2_j=\E^1_i(\G_{+j}+\G_{-j})=\G_{ij}
\end{equation}
and
\begin{equation}
\E^2_j\E^1_i=(\G_{+j}+\G_{-j})\E^1_i=\G_{ij}.
\end{equation}
A comparison of these equations proves the claim.
\end{proof}

We note that in general two observables may have many different joint observables; this fact will be demonstrated in Section \ref{sec:covjoint}.

For later use we recall the following general fact. The set of observables on a fixed outcome space is convex: for two observables $\E,\F$ and any $t\in[0,1]$, a new observable $t\E+(1-t)\F$ is defined as $\omega_i\mapsto t\E_i+(1-t)\F_i$.

\begin{proposition}\label{prop:joint-convex}
Let $(\E^1,\E^2)$ and $(\F^1,\F^2)$ be two pairs of jointly measurable observables. Then for any
$t\in[0,1]$, the observables $t\E^1+(1-t)\F^1$ and $t\E^2+(1-t)\F^2$ are jointly measurable.
\end{proposition}

\begin{proof}
Let $\G$ be a joint observable of $\E^1,\E^2$ and $\Ha$ of $\F^1,\F^2$. Then $t\G+(1-t)\Ha$ is a joint observable of $t\E^1+(1-t)\F^1$ and $t\E^2+(1-t)\F^2$.
\end{proof}

\subsection{Approximate joint measurability}\label{sec:approximate}

Assume that two observables $\A$ and $\B$ do not have a joint measurement. We may still ask if it could be possible to obtain some information on both observables in a single measurement scheme. One way of approaching this task is to consider whether there are two jointly measurable observables $\E^1,\E^2$ that are close to $\A,\B$, respectively, in a sense to be determined. Any joint measurement of $\E^1$ and $\E^2$ can then be regarded as an approximate joint measurement of $\A$ and $\B$.

A natural characterization of the closeness between two observables (assumed to have the same outcome space) is based on the degree of similarity of their associated probability distributions for all states. Hence we define the distance between observables $\A$ and $\B$ in the following way:
\begin{equation}
\D(\A,\B):=\max_{j} \sup_{T} \big| \tr[T\A_j]-\tr[T\B_j]\big|=\max_{j}\no{\A_j-\B_j}.
\end{equation}
Clearly, $0\leq\D(\A,\B)\leq 1$, and $\D(\A,\B)=0$ if and only if $\A=\B$. Moreover, the triangle inequality holds for a triple of observables, so that $\D$ is indeed a metric.
If $\A$ and $\B$ are simple observables, then $\A_--\B_-=\B_+-\A_+$ and therefore 
\begin{equation}
\D(\A,\B)=\no{\A_+ - \B_+}=\no{\A_- - \B_-}.
\end{equation}

A conventional approach to realizing approximate joint measurements consists of replacing
the observables $\A,\B$ to be approximated with some coarse-grained versions $\E^1,\E^2$. Here we briefly illustrate this approach in the case of simple observables from the perspective of the general framework. We refer to \cite[Chapter 7]{FQMEA} for a review and examples on this topic.

If $\A$ and $\B$ are simple observables, one defines, using $2\times 2$ stochastic matrices\footnote{A stochastic matrix is a square matrix whose entries are non-negative real numbers and each column sums to 1.} $(\lambda_{ik}),(\mu_{j\ell})$, the coarse-grainings $\E^1$ and $\E^2$ by
\begin{equation}\label{smear}
\E^1_i=\lambda_{i+}\A_++\lambda_{i-}\A_-,\quad \E^2_j=\mu_{j+}\B_++\mu_{j-}\B_-.
\end{equation}
We expect that $\E^1$ is a good approximation of $\A$ if $\lambda_{++}$ is close to 1 and $\lambda_{+-}$ is close to 0. Indeed, if, for instance, $\A$ is a sharp observable then
\begin{equation}\label{distance-of-coarsegraining}
\D(\E^1,\A)=\max \{ 1-\lambda_{++},\lambda_{+-}\}.
\end{equation}

To give an example of jointly measurable observables, assume that $\E^1$ and $\E^2$ are defined as in (\ref{smear}). Since we want to approximate $\A$ by $\E^1$ and $\B$ by $\E^2$, it is natural to require that
\begin{equation}\label{smatrix}
\lambda_{++}\ge\lambda_{+-},\quad \mu_{++}\ge\mu_{+-}.
\end{equation}
In fact, if these inequalities do not hold, one can choose $\lambda'_{ik}=\lambda_{-ik}$,
$\mu'_{jl}=\mu_{-jl}$ to obtain new coarse-grainings which do satisfy the inequalities.

Now, define $\G_{++}=\min\{\lambda_{+-},\mu_{+-}\}\,\id$. Condition (\ref{smatrix}) implies that $\nul\le\G_{++}\le\E^1_+$ and $\G_{++}\le\E^2_+$. The remaining inequality
required for joint measurability, $\E^1_+ + \E^2_+ - \id\le\G_{++}$, depends on the specific structure of the observables $\A$ and $\B$; it is ensured to hold independently of $\A$ and $\B$ if
\begin{equation}
\lambda_{++} + \mu_{++} \le 1 + \min\{\lambda_{+-},\mu_{+-}\}.
\end{equation}
This shows that one can always construct (nontrivial) jointly measurable coarse-grainings; a possible choice is, for instance, $\lambda_{++}=\mu_{++}=\frac{2}{3}$ and $\lambda_{+-}=\mu_{+-}=\frac{1}{3}$.

\section{Qubit observables}\label{sec:qubit}

\subsection{Effects and observables}\label{sec:qubit-effects}

In the 2-dimensional Hilbert space of a qubit one can take the unit
operator $\id$ together with the Pauli operators
$\sigma_1,\sigma_2,\sigma_3$ as a basis of the real vector space of
selfadjoint linear operators. The latter can be defined with respect
to any fixed basis of orthogonal unit vectors $\fii_+,\fii_-$ so
that the usual relations are satisfied:
$\sigma_3\fii_\pm=\pm\fii_\pm$, $\sigma_1\fii_\pm=\fii_\mp$,
$\sigma_2\fii_\pm=\pm i\fii_\mp$. We will write $\vsigma$ for the
operator triple $(\sigma_1,\sigma_2,\sigma_3)$. States of a qubit
can be written in the form $T_{\vr}=\half \left( \id +
\vr\cdot\vsigma\right)$, where $\vr\in\R^3$ and $\no{\vr}\leq 1$.
The pure states are characterized by the condition $\no{\vr}=1$.

For each $(\alpha,\va)\in\R^4$, we denote
\begin{equation}
\Aaa:=\frac{1}{2}\left( \alpha \id + \va\cdot\vsigma \right).
\end{equation}
The eigenvalues of the operator $\Aaa$ are $\frac{1}{2}(\alpha \pm
\no{\va})$. Hence, $\Aaa$ is an effect if
\begin{equation}
\no{\va}\leq \alpha \leq 2-\no{\va},
\end{equation}
which implies, in particular, that $\no{\va}\leq 1$. The operator $\Aaa$ is a nontrivial projection if
\begin{equation}
\alpha=\no{\va}=1.
\end{equation}
The spectral decomposition of the effect $\Aaa$, $\va\neq 0$, is
(putting $\hat{\va}:=\no{\va}^{-1}\va$)
\begin{equation}\label{eqn:spectraldecom}
\Aaa=\frac{1}{2}(\alpha+\no{\va})A(1,\hat\va) +
\frac{1}{2}(\alpha-\no{\va})A(1,-\hat\va).
\end{equation}
For later use we note the commutator of two effects $A(\alpha,\va)$
and $A(\beta,\vb)$:
\begin{equation}
\left[ A(\alpha,\va),A(\beta,\vb) \right]=\tfrac{1}{2}
\left(\va\times\vb \right) \cdot \vsigma.
\end{equation}

Since the Hilbert space of a qubit is 2-dimensional, any sharp qubit
observable is (effectively) simple. It is clear that this
restriction does not apply for a qubit observable in general; one
can write the identity operator $\id$ as a sum of arbitrarily many
different effects. It is also known that for some quantum
informational tasks, such as unambiguous state discrimination, one
needs other qubit observables than the simple ones; see e.g. \cite{Chefles00}. Here we shall,
however, concentrate on simple qubit observables as our aim is to
study approximate joint measurements of sharp qubit observables. 
To clarify further the nature of sharp qubit observables, we note that again due to the low dimensionality, 
the projections that constitute such an observable are of rank one, which implies that their
repeatable measurements are von Neumann measurements \cite{OQP}.

We denote by $\Eaa$ the simple qubit observable defined as
\begin{equation*}\begin{split}
\omega_+\mapsto\Eaa_{+} &:= \Aaa,\\ \omega_-\mapsto\Eaa_{-}&:=\id-\Aaa=A(2-\alpha,-\va).
\end{split}
\end{equation*}
A special case is given by the sharp observables $\E^{1,\hat\va}$.
The spectral decomposition (\ref{eqn:spectraldecom}) of $A(\alpha,\va)$ shows that $\Eaa$ is a
coarse-graining of $\Esa$.

From the above commutator formula we recover the well known fact
that the observables $\Eaa$ and $\Ebb$ commute exactly when the
vectors $\va$ and $\vb$ are collinear. Together with Proposition
\ref{prop:com}, this shows that an observable $\Ebb$ is jointly
measurable with a sharp observable $\Esa$ if and only if $\Ebb$ is a
coarse-graining of $\Esa$. A joint measurement of that kind is of
little value; one can simply measure $\Esa$ alone to get the same
information.

\subsection{Covariance}\label{sec:covariance}

Let $U$ be a unitary operator describing some symmetry
transformation of the system. We assume that $U^2=\id$, so that
$\{\id,U\}$ form a two-element group. In other words, $U$ is a
selfadjoint unitary operator. We say that an observable $\Eaa$ is
\emph{covariant with respect to} $U$, or $U$-\emph{covariant} for
short, if
\begin{equation}
U\Eaa_+U=\Eaa_-.
\end{equation}
This covariance condition means that the symmetry transformation described by $U$ swaps the outcomes of the observable $\Eaa$ but has no other effect on its measurement outcome distributions.

Effects $\Eaa_+$ and $\Eaa_-$ can be unitarily equivalent only if they have the same eigenvalues, which is the case exactly when $\alpha=1$. Hence, $\Eaa$ can be covariant only if $\alpha=1$. Assume that $\alpha=1$ and fix a unit vector $\vu\in\R^3$ orthogonal to $\va$. The operator $U=\vu\cdot\vsigma$ is a selfadjoint unitary operator and
\begin{equation}\label{eqn:Eacov}
U\Ea_+U =\Ea_-.
\end{equation}
Moreover, any selfadjoint unitary operator $U$ satisfying (\ref{eqn:Eacov}) is of the form $U=\vu\cdot\vsigma$ for some unit vector $\vu$ orthogonal to $\va$.

In \cite{AnBaAs05}, an observable $\Eaa$ was selected in relation
to a sharp observable $\En$ by the requirement that the expectation
values of $\Eaa$ are proportional to those of $\En$. This
requirement, called there {\em unbiasedness}, is equivalent with the
fact that $\alpha=1$ and the vectors $\va$ and $\vn$ are parallel.
Hence, the unbiasedness requirement means that $\Eaa$ is covariant
with respect to the same unitary operators as $\En$, i.e., the
observables $\Eaa$ and $\En$ have the same symmetry properties.

\subsection{Sharpness and distance}\label{sec:sharpness-distance}

The spectral width of an operator acting on two dimensional Hilbert
space is simply the difference of its greater and lower eigenvalues.
The sharpness of an observable $\Eaa$ is thus found to be
\begin{equation}\label{eqn:sharpness-qubit}
\S(\Eaa)=\no{\va}(1-|1-\alpha|)=\no{\va}\min\{\alpha,2-\alpha\}.
\end{equation}
With this expression one can easily confirm the statements of
Section \ref{sec:effects} for simple qubit observables: $\S(\Eaa)=1$
exactly when $\Eaa$ is a sharp observable and $\S(\Eaa)=0$ exactly
when $\Eaa$ is a trivial observable. We also note the following useful observation:
\begin{equation}\label{eqn:sharpness-ineq}
\S(\Eaa)\le\S(\Ea).
\end{equation}

The distance between two qubit
observables $\Eaa$ and $\Ebb$ is given by the formula
\begin{equation}\label{eqn:distance-ab}
\D(\Eaa,\Ebb)=\half\no{\va-\vb}+\half |\alpha-\beta|.
\end{equation}
This shows, in particular, that the distance of a given observable $\Eaa$ from any sharp observable $\En$ is minimal when $\vn=\hat\va$, or in other words, when $\Eaa$ is a coarse-graining of $\En$. In this case we have
\begin{equation}\label{eqn:distance-aa}
\D(\Eaa,\E^{1,\hat\va})=\half\left( 1-\no{\va} \right) + \half |1-\alpha|.
\end{equation}
We also note the following:
\begin{equation}
\D(\Eaa,\Esa)\ge\D(\Ea,\Esa).
\end{equation}

Finally, from equations (\ref{eqn:sharpness-qubit}) and (\ref{eqn:distance-aa}) we get the following relations:
\begin{equation}\label{eqn:dist-sharp}\begin{split}
\D(\Eaa,\En)+\half\S(\Eaa)&\ge
\D(\Eaa,\E^{1,\hat\va})+\half\S(\Eaa)\\
&\ge\D(\Ea,\E^{1,\hat\va})+\half\S(\Ea) =\half.
\end{split}
\end{equation}
The last equation shows that the distance between $\Ea$ and $\E^{1,\hat\va}$ is directly related to the sharpness of $\Ea$. This is not surprising when we recall that $\Ea$ is a coarse-graining of $\E^{1,\hat\va}$.

\section{Joint measurability of qubit observables}\label{sec:joint-qubit}

\subsection{General criterion for joint measurability}\label{sec:genjoint}

The joint measurability conditions (\ref{eqn:cone-condition}) applied to two qubit
observables $\Eaa,\Ebb$ takes the following form: there exists an
operator $\G_{++}=\half(\gamma\id+\vg\cdot\vsigma)$ such that
\begin{eqnarray}
\no{\vg}&\le&\gamma;\\
\no{\va-\vg}&\le&\alpha-\gamma;\\
\no{\vb-\vg}&\le&\beta-\gamma;\\
\no{\va+\vb-\vg}&\le&2+\gamma-\alpha-\beta.
\end{eqnarray}
Let $\bB(\vx,r)$ denote the closed ball with center $\vx$ and radius
$r$. Then it is seen that the joint measurability of $\Eaa,\Ebb$ is
equivalent to the statement that there exists a number $\gamma\ge 0$ such
that the intersection of four balls is non-empty:
\begin{equation}\label{eqn:ball-coex}
\bB(\vnull,\gamma)\,\cap\, \bB(\va,\alpha-\gamma)\,\cap\,\bB(\vb,\beta-\gamma)\,\cap\,\bB(\va+\vb,2+\gamma-\alpha-\beta)\ne\emptyset.
\end{equation}

The criterion (\ref{eqn:ball-coex}) immediately gives the
following as a necessary condition for joint measurability: the two pairs of balls diagonally opposite to each
other must have separations which are no greater than the sum of
their radii; thus, there must be a $\gamma\ge 0$ such that
\begin{eqnarray}
\no{\va-\vb}&\le&\alpha+\beta-2\gamma,\\
\no{\va+\vb}&\le&2-\alpha-\beta+2\gamma,
\end{eqnarray}
or equivalently,
\begin{equation}\label{eqn:gamma12}
\gamma_1:=\half\no{\va+\vb}+\half[\alpha+\beta-2]\le
\gamma\le\half[\alpha+\beta]-\half\no{\va-\vb}=:\gamma_2.
\end{equation}
This gives an interval for $\gamma$ to lie in which has to be
nonempty. Therefore the following is a necessary joint measurability
condition:
\begin{equation}
\gamma_2-\gamma_1=1-[\half\no{\va+\vb}+\half\no{\va-\vb}]\ge0.
\end{equation}

\begin{proposition}\label{prop:generalineq}
If observables $\Eaa$ and $\Ebb$ are jointly measurable,
then\footnote{This condition has the following geometric meaning:
for a observable $\Ea$, a jointly measurable observable $\Eb$ is
such that the vector $\vb$ is inside a prolate spheroid. The center
of the spheroid is in the origin and its major axis is in the
direction of $\va$. The polar radius of the spheroid is 1 and the
equatorial radius is $(1-\no{\va}^2)^{1/2}$. In fact, in coordinates
for which $\va$ is in the $z$-direction, the inequality  becomes
$b_x^2+b_y^2+(1-a^2)b_z^2\le1-a^2$, to be read as a condition for
$\vb$.}
\begin{equation}\label{eqn:generalineq}
\no{\va + \vb} + \no{\va - \vb} \leq 2.
\end{equation}
\end{proposition}

In the case of covariant qubit observables (for which $\alpha=\beta=1$) the condition (\ref{eqn:generalineq}) is found to be also sufficient for joint measurability, as was
shown in \cite{Busch86}. A new proof of this fact, stated below, will arise as a corollary of our investigation in Subsection \ref{sec:covjoint}.

\begin{proposition}\label{prop:Paulsineq}
Observables $\Ea$ and $\Eb$ are jointly measurable if and only if
inequality $(\ref{eqn:generalineq})$ holds.
\end{proposition}

In the following example we demonstrate that (\ref{eqn:generalineq})
is not sufficient in general to guarantee the joint measurability of
observables $\Eaa$ and $\Ebb$.

\begin{example}
Let us consider the case where the vectors $\va$ and $\vb$ are
orthogonal and  equality holds in (\ref{eqn:generalineq}), or in
other words, $\no{\va+\vb}=\no{\va-\vb}=1$. Assume that $\Eaa$ and
$\Ebb$ are jointly measurable observables. We have
$\gamma=\gamma_1=\gamma_2$ and therefore, there is only one point
$\vg$ in the intersection
$\bB(\va,\alpha-\gamma)\cap\bB(\vb,\beta-\gamma)$, and similarly in
the intersection
$\bB(\vnull,\gamma)\cap\bB(\va+\vb,\gamma+2-\alpha-\beta)$.
Thus, $\vg$ is in the boundary of $\bB(\va,\alpha-\gamma)$ and it must satisfy the equation
\begin{equation}
\vg = \va+(\alpha-\gamma)(\vb-\va)
\end{equation}
and three similar equations corresponding to the other balls. These
equations taken together imply that $\alpha=\beta=1$. As condition
(\ref{eqn:generalineq}) does not restrict $\alpha$ and $\beta$, we
conclude that (\ref{eqn:generalineq}) is not sufficient to ensure
the joint measurability of $\Eaa$ and $\Ebb$. In fact, we could have
chosen $\alpha=\beta=\no{\va}=\no{\vb}=1/\surd 2$, in which case the
observables $\Eaa$ and $\Ebb$ are not jointly measurable although
(\ref{eqn:generalineq}) is satisfied.
\end{example}

Propositions \ref{prop:generalineq} and \ref{prop:Paulsineq} lead to the following observation, which we will need later.

\begin{proposition}\label{prop:alsojoint}
If $\Eaa$ and $\Ebb$ are jointly measurable, then also $\Ea$ and
$\Eb$ are jointly measurable.
\end{proposition}

\subsection{Sufficient conditions for joint measurability}\label{sec:sufficient}

The problem of finding necessary {\em and} sufficient conditions for the joint measurability of a pair of qubit observables $\Eaa$ and $\Ebb$ beyond the above case of $\Ea,\Eb$ has only recently been solved by the present authors in different collaborations. In \cite{StReHe08}, this is achieved by analyzing the sphere intersection condition (\ref{eqn:ball-coex}), whereas in \cite{BuSch08} the cone intersection condition (\ref{eqn:cone-condition}) is elucidated. The sets of inequalities found for $\alpha,\va,\beta,\vb$ are rather involved and not easily comparable, hence we refrain from reproducing them here.
Instead we give a  sufficient condition for the joint measurability of $\Eaa$ and
$\Ebb$ which is an obvious strengthening of (\ref{eqn:generalineq}). The fact that this stronger
condition may appear quite natural at first sight but is actually not necessary highlights the intricate nature of the general problem solved in \cite{StReHe08} and \cite{BuSch08}. 

First we identify two distinguished effects $A_1:=A(\gamma_1,\vg_1)$ and
$A_2:=A(\gamma_2,\vg_2)$, where $\gamma_1,\gamma_2$ are the parameters from
Eq.~(\ref{eqn:gamma12}) and
\begin{equation}\begin{split}
\vg_1&=
\frac 12\left[ 1-\frac{2-\alpha-\beta}{\no{\va+\vb }} \right]\,( \va+\vb ),\\
\vg_2&
=\frac 12(\va+\vb )-\frac{\alpha-\beta}{\no{\va-\vb }}\frac 12
(\va-\vb)  .
\end{split}
\end{equation}
The effect $A_1$ is in $\cC^\lor(\nul)\cap\cC^\lor(\E^1_++\E^2_+-\id)$ and it is the unique
element of all effects $A(\gamma,\vg)$ in that intersection with the lowest possible
$\gamma$. The effect $A_2$ is in $\cC^\land(\E^1_+)\cap\cC^\land(\E^2_+)$
and it is the unique element of all effects $A(\gamma,\vg)$ in that intersection with the
greatest possible $\gamma$. Now, joint measurability is guaranteed if $A_1\le A_2$,
which is equivalent to the condition:
\begin{equation}\label{eqn:sufficient}
\no{\va+\vb}+\no{\va-\vb}
+\no{\frac{2-\alpha-\beta}{\no{\va+\vb}}(\va+\vb)-
\frac{\alpha-\beta}{\no{\va-\vb}}(\va-\vb)}
\le 2.
\end{equation}
It is not hard to verify that  this condition is automatically satisfied in all trivial and commutative cases,
where joint measurability is given. Furthermore it follows from the stronger condition
\begin{equation}\label{eqn:strong-suff}
\no{\va+\vb}+\no{\va-\vb}+|2-\alpha-\beta|+ |\alpha-\beta| \le 2,
\end{equation}
which can also be written in operator terms as
\begin{equation}
\no{\Eaa_+-\Ebb_+}+\no{\Eaa_+-\Ebb_-}\le 1.
\end{equation}
This sufficient condition for joint measurability is satisfied in all cases with $\alpha=\beta=1$ but is generally not necessary, as can be seen from the example $\Eaa_+=\id$, $\Ebb_+=A(1,\vn)$ (where
$\vn$ is any unit vector).

In nontrivial cases the above sufficient joint measurability conditions can be further strengthened and simplified. For two qubit observables $\Eaa$ and $\Ebb$, the nontriviality requirement in the sense of Example \ref{ex:trivial-coex} amounts to the following:
\begin{equation}\label{eqn:nontrivial}\begin{split}
|\alpha-\beta|&< \no{\va-\vb}\quad (\mathrm{not\ (a),(b)}); \\
|2-\alpha-\beta|&< \no{\va+\vb}\quad (\mathrm{not\ (c),(d)}).
\end{split}
\end{equation}
Under these nontriviality assumptions, the conditions (\ref{eqn:sufficient}) and (\ref{eqn:strong-suff}) are
seen to be satisfied if
\begin{equation}
\no{\va+\vb}+\no{\va-\vb}\le 1.
\end{equation}

The next example shows that the sufficient condition
(\ref{eqn:sufficient}) is not a necessary condition.
\begin{example}
We consider the case where $\va\perp\vb$. Furthermore, let $\va=\hat\va$ be a unit vector, so that
$A:=\Eaa_+=\alpha A(1,\hat\va)$ is a multiple of a projection. Note that $A$ being an effect entails that
$\alpha\le 1$. Next we denote $B:=\Ebb_+=A(1,\vb)$, where we assume that $b:=\no{\vb}\ne 0$. 

Joint measurability of $A,B$ is given if and only if there is an operator $G$ which is bounded above by $A,\ B$ and bounded below by $\nul,\ A+B-\id$. The inequality $G\le A$ is satisfied if and only if $G$ is a multiple of the projection $A(1,\hat\va)$, hence: $G=\gamma A(1,\hat\va)$, and $\gamma\le\alpha$. Further, $\gamma$
must be chosen such that $\gamma A(1,\hat\va)\le B$; thus:
\[
1-\gamma\ge\sqrt{\gamma^2+b^2}.
\]
This is equivalent to $\gamma\le\gamma_0:=\half(1-b^2)$.
The inequality $\id-A-B+G\ge\nul$ is equivalent to
\[
1-\alpha+\gamma\ge \sqrt{(\alpha-\gamma)^2+b^2}.
\]
This is solved by $\gamma\ge \alpha-\half(1-b^2)=\alpha-\gamma_0$. 

To summarize: the given effects $A,B$ are jointly measurable if and only if
\[
\alpha-\gamma_0\le\min\{\gamma_0,\alpha\},\quad \gamma_0\equiv\half(1-b^2).
\]
The nontriviality conditions assume here the form
\[
|\alpha-\beta|=|2-\alpha-\beta|=1-\alpha < \no{\va-\vb}=\no{\va+\vb}=\sqrt{\alpha^2+b^2},
\]
which is equivalent to $\gamma_0<\alpha$. In this case the joint measurability condition reduces to
$\alpha/2\le\gamma_0$.

We are now ready to show that condition (\ref{eqn:sufficient}) can be violated in nontrivial cases. In the given constellation, this inequality assumes the form
\[
\sqrt{\alpha^2+b^2}+\frac{\alpha(1-\alpha)}{\sqrt{\alpha^2+b^2}}\le 1.
\]
For the choice $\alpha=\half=1-b^2=2\gamma_0$, the left hand side becomes $2/\sqrt 3$, which is greater than 1. However, this choice fulfills the joint measurability and nontriviality conditions.
\end{example}

\subsection{Covariant joint observables}\label{sec:covjoint}

In what follows we will investigate implications of covariance. In
this way we establish a far-reaching analogy to similar studies made
on approximate joint measurements of position and momentum where
covariance (under translations on phase space) was found to be
paramount (cf. the review \cite{BuHeLa06}).

Let us first note that there is a unitary operator $U$ such that
both $\Ea$ and $\Eb$ are covariant with respect to $U$. Namely, fix
a unit vector $\mathbf{u}$ orthogonal to both $\mathbf{a}$ and
$\mathbf{b}$ and choose $U=\vu\cdot\vsigma$.

We say that a joint observable $\G$ of $\Ea$ and $\Eb$ is covariant
with respect to $U$, or $U$-covariant, if
\begin{equation}\label{covG}
\begin{array}{ll}
U\G_{++}U=\G_{--}, & \\
U\G_{+-}U=\G_{-+}. &
\end{array}
\end{equation}
Since $\Ea$ and $\Eb$ are $U$-covariant, the two equations in
(\ref{covG}) are equivalent and thus, already one of them implies
that $\G$ is $U$-covariant.

\begin{proposition}\label{prop:covjoint}
If $\Ea$ and $\Eb$ are jointly measurable, then they have a
$U$-covariant joint observable.
\end{proposition}

\begin{proof}
Let $\G$ be a joint observable of $\Ea$ and $\Eb$. Define
\begin{eqnarray*}
\widetilde{\G}_{++} &=& \half \left( \G_{++} + U\G_{--}U \right), \\
\widetilde{\G}_{+-} &=& \half \left( \G_{+-} + U\G_{-+}U \right), \\
\widetilde{\G}_{-+} &=& \half \left( \G_{-+} + U\G_{+-}U \right), \\
\widetilde{\G}_{--} &=& \half \left( \G_{--} + U\G_{++}U \right).
\end{eqnarray*}
Each operator $\widetilde{\G}_{\pm\pm}$ is a convex combination of
two effects, hence an effect. Moreover, the sum of these effects is
$\id$ and thus, $\widetilde{\G}$ is an observable.

We have
\begin{eqnarray*}
\widetilde{\G}_{++} + \widetilde{\G}_{+-} &=& \Ea_+,\\
\widetilde{\G}_{++} + \widetilde{\G}_{-+} &=&\Eb_+,
\end{eqnarray*}
showing that $\widetilde{\G}$ is a joint observable of $\Ea$ and $\Eb$.
Using the fact that $U^2=\id$ we immediately see that $U\widetilde{\G}_{++}U=\widetilde{\G}_{--}$, meaning that $\widetilde{\G}$ is $U$-covariant.
\end{proof}

We proceed by characterizing all $U$-covariant joint observables of
$\Ea$ and $\Eb$. Denoting $\G_{++}=\half \left( \gamma\id +
\vg\cdot\vsigma \right)$ the covariance
condition (\ref{covG}) can be written in the form
\begin{equation}
\vg-(\vu\cdot\vg)\vu=\half(\va+\vb),
\end{equation}
which means that $\vg=\half(\va+\vb)+p\vu$ for some $p\in\R$.  The joint measurability condition (\ref{simple-coex}) reduces to the requirement that
\begin{equation}\label{gammap}
\sqrt{\tfrac{1}{4}\no{\va+\vb}^2+p^2} \leq \gamma \leq 1 -\sqrt{\tfrac{1}{4}\no{\va-\vb}^2+p^2}.
\end{equation}
We conclude that $U$-covariant joint observables of $\Ea$ and $\Eb$
are characterized by the pairs $(\gamma,p)$ satisfying
(\ref{gammap}). The covariant joint observable $\G$ corresponding to $(\gamma,p)$ is
\begin{eqnarray*}
\G_{++} &=& \frac{\gamma}{2} \id + \frac{1}{4}(\va+\vb)\cdot\vsigma+\frac{p}{2}\vu\cdot\vsigma, \\
\G_{+-} &=& \frac{1-\gamma}{2} \id + \frac{1}{4}(\va-\vb)\cdot\vsigma - \frac{p}{2}\vu\cdot\vsigma, \\
\G_{-+} &=& \frac{1-\gamma}{2} \id - \frac{1}{4}(\va-\vb)\cdot\vsigma - \frac{p}{2}\vu\cdot\vsigma, \\
\G_{--} &=& \frac{\gamma}{2} \id -\frac{1}{4}(\va+\vb)\cdot\vsigma + \frac{p}{2}\vu\cdot\vsigma.
\end{eqnarray*}

If a pair $(\gamma,p)$ satisfies condition (\ref{gammap}), then so
does $(\gamma,0)$. Hence, $\Ea$ and $\Eb$ have a $U$-covariant joint
observable if and only if there is a $\gamma$ such that
\begin{equation}\label{gamma}
\half\no{\mathbf{a}+\mathbf{b}} \leq \gamma \leq 1 -
\half\no{\mathbf{a}-\mathbf{b}},
\end{equation}
or equivalently, if and only if inequality (\ref{eqn:generalineq})
holds. This together with Proposition \ref{prop:covjoint} gives the result
cited in Proposition \ref{prop:Paulsineq}. Inequality (\ref{gamma}) implies that
\begin{equation}\label{eqn:covariantineq}
\half\no{\mathbf{a}+\mathbf{b}} \leq \half( 1+\va\cdot\vb )
\leq 1- \half\no{\mathbf{a}-\mathbf{b}}.
\end{equation}
Thus, if $\Ea$ and $\Eb$ are jointly measurable, then they have a joint observable $\G^0$ corresponding to the choice $\gamma=\gamma_0:=\half( 1+\va\cdot\vb )$ and $p=0$. The effects of $\G^0$ can be written in the form
\begin{equation}
\G^0_{ij}=\half \left( \Ea_i\Eb_j + \Eb_j\Ea_i \right),\quad i,j=\pm.
\end{equation}

If we have the limiting case of condition (\ref{gamma}), i.e. 
\begin{equation}
\half\no{\mathbf{a}+\mathbf{b}}=1 -
\half\no{\mathbf{a}-\mathbf{b}},
\end{equation}
then $(\gamma_0,0)$ is the only possible pair
and hence, in this case $\G^0$ is the unique $U$-covariant joint observable of $\Ea$ and $\Eb$.
In all other situations of covariant joint measurements except this limiting case, there is a
continuum of possible pairs $(\gamma,p)$. The joint observable $\G$
corresponding to $(\gamma,p)$ is informationally complete if and
only if $p\neq 0$; this follows directly from \cite[Theorem
4.7]{Busch86}.

Finally, we note that the covariance of $\Ea$ and $\Eb$ does not imply that they have only covariant joint observables. To give an example, assume that the vectors $\va$ and $\vb$ satisfy $\va\cdot\vb\geq 0$ and $\no{\va+\vb} < 1$, so that $\Ea$ and $\Eb$ are jointly measurable. Fix a number $t$ such that $0<t\leq \no{\va+\vb}^{-1} -1$, and define
\begin{eqnarray*}
\G_{++} &=& \frac{1}{4} \id + \frac{1}{4}(1+t)(\va+\vb)\cdot\vsigma, \\
\G_{+-} &=& \frac{1}{4} \id + \frac{1}{4}((1-t)\va-(1+t)\vb)\cdot\vsigma, \\
\G_{-+} &=& \frac{1}{4} \id + \frac{1}{4}((1-t)\vb-(1+t)\va)\cdot\vsigma \\
\G_{--} &=& \frac{1}{4} \id -\frac{1}{4}(1-t)(\va+\vb)\cdot\vsigma.
\end{eqnarray*}
Then $\G$ is a joint observable for $\Ea$ and $\Eb$ but it is not covariant.
Indeed, the above condition on $t$ guarantees that inequalities (\ref{simple-coex}) are satisfied. The eigenvalues of $\G_{++}$ and $\G_{--}$ are different and thus, the covariance condition (\ref{covG}) cannot be satisfied with any unitary operator $U$.

\subsection{Joint measurability vs. sharpness}\label{sec:jm-sharpness}

It is instructive to write down the joint measurement condition for two covariant observables $\Ea$ and $\Eb$, assuming that the vectors $\va$ and $\vb$ are orthogonal. The inequality (\ref{eqn:generalineq}) takes now the form
\begin{equation}
\U(\Ea)+\U(\Eb)\geq 1,
\end{equation}
showing that the joint measurability is achieved exactly when the observables are made unsharp enough. 

In the general case, we can transform (\ref{eqn:generalineq}) by repeated squaring into the equivalent inequality
\begin{equation}\label{eqn:generalineq2}
\no{\va}^2 + \no{\vb}^2 \leq 1+ (\va\cdot\vb)^2,
\end{equation}
which can be written in the form
\begin{equation}
\no{\va\times\vb}^2\le(1-\no{\va}^2)(1-\no{\vb}^2).
\end{equation}
The term on the left hand side is equal to $4\no{[\Eaa_+,\Ebb_+]}^2$.
The term on the right hand side turns out to give a bound for the degrees
of sharpness of $\Eaa$ and $\Ebb$. Considering the formulas (\ref{eqn:u}) and (\ref{eqn:sharpness-ineq})
we obtain the following.
\begin{proposition}
If two qubit observables $\Eaa,\Ebb$ are jointly measurable,
the degrees of their unsharpness satisfy the inequality
\begin{equation}\label{eqn:jm-unsharp}
\U(\Eaa)\,\U(\Ebb)\,\ge\,\U(\Ea)\,\U(\Eb)\,\ge \,4\no{[\Eaa_+,\Ebb_+]}^2.
\end{equation}
If $\alpha=\beta=1$, this inequality is in fact equivalent to the joint measurability condition.
\end{proposition}
This shows that the intrinsic sharpness of two jointly measurable simple qubit observables
$\Eaa$ and $\Ebb$ is limited by the noncommutativity of the generating effects.

\section{Approximate joint measurement for two sharp qubit observables}\label{sec:approx-joint}

Two sharp observables $\En$ and $\Em$ are jointly measurable exactly
when they commute, and this happens if and only if $\vn=\pm\vm$. In
this section we consider the case $\vn\neq\pm\vm$, so that only
approximate joint measurements are possible. The idea is to choose a
jointly measurable pair $(\Eaa,\Ebb)$ to approximate the sharp pair
$(\En,\Em)$. To be specific, and without loss of generality, we assume
$\cos\theta:=\vn\cdot\vm > 0$.

We call a point $\left(\D_1,\D_2\right)\in[0,1]\times[0,1]$
\emph{admissible} if  $\D_1=\D(\Eaa,\En)$ and $\D_2=\D(\Ebb,\Em)$ for some jointly measurable observables $\Eaa$ and $\Ebb$. Not all points in the square $[0,1]\times[0,1]$ are admissible; for instance the point $(0,0)$ is not an admissible point since this would mean that $\Eaa=\En$ and $\Ebb=\Em$. We show in the following that there are also other points which are not admissible. The set of admissible points gives us a characterization on the quality of possible approximate joint measurements. 

The search for admissible points $\left( \D_1,\D_2\right)$
is narrowed down by the following simple observation:
\begin{example}
Let $\alpha\in[0,2]$. 
Then $\D(\E^{\alpha,\vnull},\En)=\half\max\{\alpha,2-\alpha\}$ and therefore
\begin{eqnarray}
\left\{\D(\E^{\alpha,\vnull},\En)\,:\,\alpha\in[0,2]\right\}
=[\half,1].
\end{eqnarray}
\end{example}

Thus, approximations by means of trivial observables
will never give distances below $\half$. Furthermore, since
$\E^{\alpha,\vnull}$ is jointly measurable with any observable
$\Ebb$, and since $\D(\Ebb,\Em)$ can assume any value in $[0,1]$, it
follows that all points in the set
$[0,1]\times[0,1]\setminus[0,\half]\times[0,\half]$ are trivially admissible.
We will therefore concentrate on admissible points $\left(\D_1,\D_2\right)$ in the region $[0,\half]\times[0,\half]$.

The next two results are not complicated but require some preparation and will be proven in the Appendix.
\begin{proposition}\label{prop:admiss-realization}
Any admissible point
$\left(\D_1,\D_2\right)\in[0,\half]\times[0,\half]$ has a
realization of the type $\D_1=\D(\Ea,\En)$, $\D_2= \D(\Eb,\Em)$,
where $\va$ and $\vb$ are in the plane spanned by $\vn$ and $\vm$.
\end{proposition}
\begin{proposition}\label{prop:convex}
The set of admissible points is a closed convex set which is reflection
symmetric with respect to the axis $\D_1=\D_2$; that is, with every
admissible point $\left(\D_1,\D_2\right)$ the point $\left(\D_2,\D_1\right)$
is also admissible. Thus the segment of the boundary curve defined as
the graph of the function
\begin{equation}
\D_1\mapsto\inf\{\D_2\,:\,\left(\D_1,\D_2\right)\ \mathrm{is\ admissible}\}
\end{equation}
is convex, symmetric and belongs to the set of admissible points.
\end{proposition}

\begin{example}\label{ex:coex0}
If $\D_1=\D(\Ea,\En)=0$ (i.e. $\va=\vn$), then the joint measurability
requirement implies that $\va||\vb$ and thus,
$$
\D(\Eb,\Em)=\half\no{\vb-\vm}\geq \half \sqrt{1-(\vn\cdot\vm)^2}=\half\sin\theta.
$$
The lower bound is attained when $\vb=\cos\theta\ \vn=(\vn\cdot\vm)\vn$.
We conclude that
$\left( 0,\half\sin\theta \right)$ and $\left( \half\sin\theta,0 \right)$
are points in the boundary of the admissible region.
\end{example}

\begin{figure}\label{fig:admissible}
\begin{center}
\includegraphics[width=7cm]{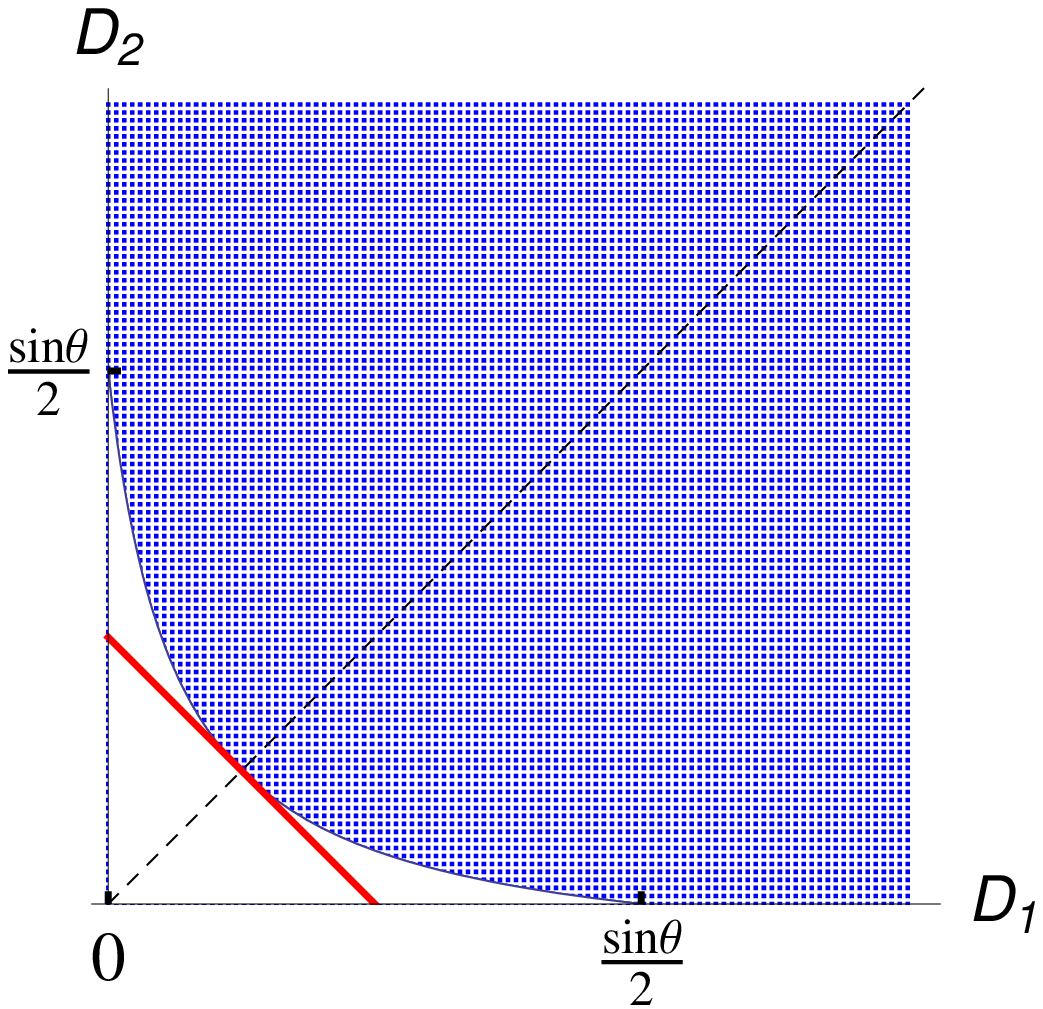}
\caption{The admissible region (dotted area) and the line $\D_1+\D_2=2\D_0$ (thick line). The dashed line is the symmetry axis $\D_1=\D_2$.}
\end{center}
\end{figure}

We next determine the boundary point with $\D_1=\D_2=:\D_0$. Due to the convexity
of the admissible region and its reflection symmetry  with respect to the line $\D_1=\D_2$,
it follows immediately that the admissible region is bounded below tightly by the straight line
$\D_1+\D_2=2\D_0$. This situation is sketched in Figure \ref{fig:admissible}. Determination of the value of $\D_0$ yields the following result.

\begin{proposition}\label{prop:D0}
 Any admissible point $\left(\D_1,\D_2\right)=\left(\D(\Eaa,\En),\D(\Ebb,\Em)\right)$ satisfies
 the inaccuracy trade-off relation
\begin{equation}\label{eqn:dist-approx}
\D(\Eaa,\En)+\D(\Ebb,\Em)\ge 2\D_0,
\end{equation}
where
\begin{equation}
2\D_0=\tfrac1{\sqrt2}\left[\half\no{\vn+\vm}+\half\no{\vn-\vm}-1\right]=
\tfrac1{\sqrt2}\left(\cos\tfrac\theta2+\sin\tfrac\theta2-1\right).
\end{equation}
The point $\left(\D_0,\D_0\right)$ is admissible.
\end{proposition}

\begin{proof}
Consider the set of all jointly measurable covariant observables $\Ea,\Eb$
such that $\va,\vb$ have equal fixed distance from $\vn,\vm$,
respectively: $\no{\va-\vn}=\no{\vb-\vm}\equiv d$ (so that
$\D(\Ea,\En)=\D(\Eb,\Em)=d/2$). If $(\va,\vb)$ is not symmetric
under reflection with respect to the line parallel to $\vn+\vm$, denote by
$\bar{\va}$ and $\bar{\vb}$ the mirror images of $\vb$ and $\va$,
respectively. Then, if $\Ea,\Eb$ are jointly measurable, so are
$\E^{1,\bar{\va}},\E^{1,\bar{\vb}}$ as the condition (\ref{eqn:generalineq}) is invariant under reflections. Due to Proposition \ref{prop:joint-convex}, the observables
$\half\Ea+\half\E^{1,\bar{\va}}=\E^{1,\half(\va+\bar{\va})}$ and
$\half\E^{1,\vb}+\half\E^{1,\bar{\vb}}=\E^{1,\half(\vb+\bar{\vb})}$ are jointly measurable. It is clear from their definitions that the vectors $\half(\va+\bar{\va})$ and $\half(\vb+\bar{\vb})$ are mirror images of each other. As $\va,\vb$ have equal distance $d$ from $\vn,\vm$, respectively, this means that $\va$ and $\bar{\va}$ have equal distance $d$ from $\vn$. It follows that the distance from $\vn$ to $\half(\va+\bar{\va})$ is less than $d$ (or $d$ if $\va=\bar{\va}$). We conclude that if $\va,\vb$ are not mirror images of each other, there is a pair of jointly measurable covariant observables with smaller (equal) distances from $\En,\Em$ and mirror symmetric vectors. This shows that the minimal equal distance approximations of $\En,\Em$ by means of jointly measurable
observables occur among the covariant pairs with $\va,\vb$ mirror symmetric with respect to $\vn+\vm$.

If coordinates are chosen such that
$\vn=(\sin\frac\theta2,\cos\frac\theta2)$,
$\vm=(-\sin\frac\theta2,\cos\frac\theta2)$, then let a symmetric
pair $\va,\vb$ be given by $\va=(u,v)$ and $\vb=(-u,v)$, with
$u,v>0$. For such pairs, the joint measurability condition for
$\Ea,\Eb$ assumes the form $u+v\le1$. It follows that the shortest
(equal) distances $d$ of $\va,\vb$ from $\vn,\vm$ are assumed when
$u+v=1$ and $\vn-\va$ is perpendicular to the line $u+v=1$. But this
distance $d$ is equal to the distance of the lines $u+v=1$ and
$u+v=\cos(\frac\theta2)+\sin(\frac\theta2)$, hence
\[
d=\tfrac1{\sqrt2}\left(\cos\tfrac\theta2+\sin\tfrac\theta2-1\right).
\]
\end{proof}

\begin{figure}\label{fig:arrows}
\begin{center}
\includegraphics[width=8cm]{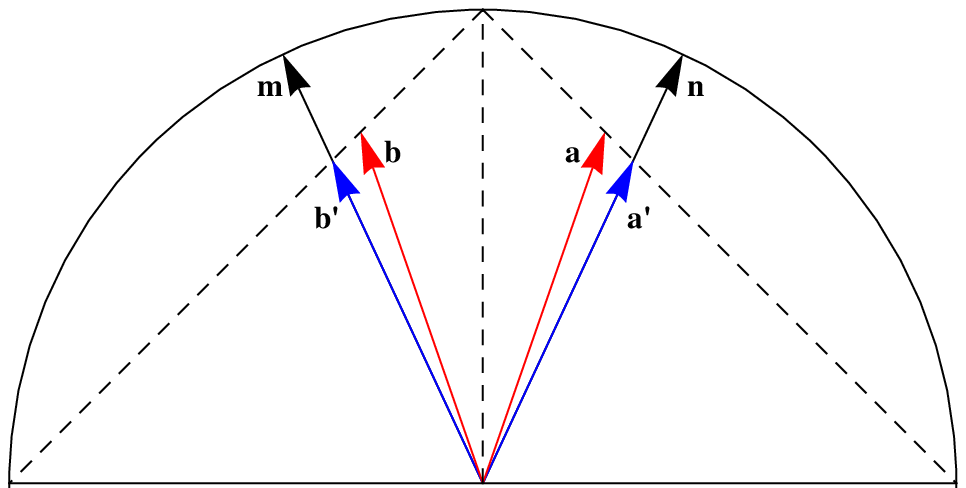}
\caption{The vectors corresponding to the optimal approximations $\Ea,\Eb$ and of optimal coarse-grainings $\Ead,\Ebd$.}
\end{center}
\end{figure}

The result of Proposition \ref{prop:D0} shows in which way the quality of the approximations is limited by the separation of the sharp observables to be  approximated in a simultaneous measurement. This relation becomes perhaps even more transparent when we write the number $\D_0$ in the form
\begin{equation}\label{eqn:D0}
\D_0=\tfrac1{2\sqrt 2}\left[\D(\En,\Em)+\D(\En,\Emn)-1\right].
\end{equation}
The appearance of $\D(\En,\Emn)$ in (\ref{eqn:D0}) is explained by the fact that the joint measurability criterion is blind to the labeling of outcomes. 

Note that 
$\cos\frac\theta2+\sin\frac\theta2=(1+\sin\theta)^{1/2}$ and $\sin\theta=\no{\vn\times\vm}=2\no{[\En,\Em]}$. Thus $\D_0$ is an increasing function of the degree of noncommutativity of the sharp observables to be estimated.

The approximations $\Ea$ and $\Eb$ leading to the boundary point $(\D_0,\D_0)$ are generally not among the  coarse-grainings of $\En$ and $\Em$ (in the sense of Section \ref{sec:approximate}). Indeed, let us denote by $\D_0^c$ the smallest number achieved under the assumptions that $\D_0^c=\D(\Ead,\En)=\D(\Ebd,\Em)$ and that $\Ead,\Ebd$ are jointly measurable and coarse-grainings of $\En,\Em$, respectively. If the vectors $\vn$ and $\vm$ are orthogonal, then $\D_0^c=\D_0$. However, if $0<\theta<\frac{\pi}{2}$, then
\begin{equation}
\D_0^c=\half \left( 1- \frac{\sqrt{1-\sin\theta}}{\cos\theta} \right) > \D_0.
\end{equation}
The vectors $\va,\vb$ and $\va',\vb'$ are illustrated in Figure \ref{fig:arrows}.
We conclude that to attain the best jointly measurable approximations of two sharp qubit observables, we are forced to seek approximating observables beyond their coarse-grainings.  

Finally, we note that it would be interesting to determine the full convex boundary curve of the region of admissible points $\left(\D_1,\D_2\right)$. Some numerically calculated boundary curves are drawn in \cite{HeReSt08}, but their analytic form is not yet known.

\section{Conclusion and outlook}\label{sec:conclusion}

In this paper we have quantified the necessary inaccuracies in approximating noncommuting sharp qubit observables by means of a pair of jointly measurable pair of observables (Eq.~(\ref{eqn:dist-approx})). We also exhibited the necessary unsharpness that observables $\Eaa,\Ebb$ must have in order to be jointly measurable (Eq.~(\ref{eqn:jm-unsharp})).
If a sharp observable $\E^{1,\hat \va}$ is approximated by one of its coarse-grainings $\Eaa$, the distance is related to the sharpness of $\Eaa$ via the relation (\ref{eqn:dist-sharp}).

Trough the case study of qubit observables we have demonstrated the conceptual difference of measurement inaccuracy and intrinsic unsharpness. This sheds some new light on the joint measurement problem raised by Uffink in \cite{Uffink94}, so we shortly recall his argumentation. Uffink analyzed a definition of ``non-ideal" or ``unsharp" joint measurement of two noncommuting observables that had previously been sketched out more or less informally by various authors. This definition
captures the idea that smearings of two noncommuting sharp observables may have a
joint observable. As it was formulated, this definition allowed
any smeared or coarse-grained version of an observable to be an approximation of
that observable, without further stipulations on the quality of the approximation. This entails
that even trivial observables (which are always among the coarse-grainings of any observable)
can be taken to represent a sort of non-ideal measurement of a given observable.

Uffink presented an example that makes this definition look absurdly comprehensive and indeed counter-intuitive: he considered two pairs of observables, $(\sigma_x,Q)$ and $(\sigma_z,P)$ and took
$\sigma_x$ as a coarse-graining of the first pair and $P$ as a coarse-graining of the second. Then
$(\sigma_x,P)$ is a joint observable for these two, and according to the letter of the
definition, it would have to be considered as representing a non-ideal or unsharp joint measurement of the original pairs.

Now, Uffink argued that while the final joint observable had $\sigma_x$ and $P$ as coarse-grainings (namely, marginals), the original observables were in no way coarse-grainings of it. Hence there was no plausible sense in which $(\sigma_x,P)$ could be regarded as representing a non-ideal joint measurement of the original pairs of observables. He thus pointed out rightly that a universal definition or criterion of approximate joint measurability was missing. But then he jumped to the conclusion  that POMs  do not contribute to solving the joint measurement problem.

We think that the present paper and many preceding it demonstrate that POMs do provide an appropriate language to clarify the definition and quantification of approximate measurements, and to determine any limitations to the accuracy of joint approximations of noncommuting pairs of observables. It is obvious that any measurement can be considered as an ``approximate" joint
measurement of an arbitrary collection of observables. Even doing nothing and
randomly picking outcomes constitutes a trivial ``non-ideal" joint
measurement of any given set of observables. There is no problem in
allowing a definition of non-ideal or approximate joint
measurements to include trivial cases; what makes any such
definition useful is whether it allows one to give quantifications
of how well each of the observables in question is being
approximated by a given scheme. As we have shown in this paper and its companion
\cite{BuHeLa06}, such quantifications
can indeed be formulated and yield a nontrivial notion of approximate measurement, leading
to the conclusion that there are universal limitations to the accuracies with which noncommuting pairs of observables can be approximately measured together. 

If the quality of the approximation is to be  optimized, the approximating observables being measured jointly must be unsharp; and the required
degree of unsharpness is linked with the quality of the approximations specified.
Using the definition of approximation introduced here, and keeping in mind the conceptual difference between the relation of approximation and the property of intrinsic unsharpness, it is clear that the above ``absurd" example considered by Uffink is simply not based on good approximations and would therefore not be regarded as a useful joint measurement.

The quantifications of inaccuracy and intrinsic unsharpness presented here for the case of qubit observables complements analogous investigations carried out in the case of continuous observables in \cite{BuHeLa06, BuPe06, CaHeTo07}. A unified approach and associated trade-off relations for the approximate joint measurements of general pairs of noncommuting quantities is still outstanding.

\section*{Appendix: Proofs of Propositions \ref{prop:admiss-realization} and \ref{prop:convex}}
%\addcontentsline{toc}{section}{Appendix: Proofs of Propositions \ref{prop:admiss-realization} and \ref{prop:convex}}

%We begin with some preparatory observations.\\

\noindent
(a) If $(\D_1,\D_2)$ is an admissible point, then also $(\D_2,\D_1)$
is an admissible point.\\
{\em Proof.}
If $(\alpha,\va)$ and $(\beta,\vb)$ realize the distances $\D_1$ and $\D_2$,
respectively, then choose $(\alpha',\va')$ and $(\beta',\vb')$ as follows: $\alpha'=\beta$,
$\va'$ has the length of $\vb$ and its angle relative to $\vn$ is equal to the angle between
$\vb$ and $\vm$; similarly, $\beta'=\alpha$, $\vb'$ has the length of $\va$ and its angle relative to
$\vm$ is the same as the angle between $\va$ and $\vn$. This ensures that $(\D_1',\D_2')=
(\D_2,\D_1)$.
\qed\\

\noindent (b) Assume that $\left( \D_1,\D_2
\right)=\left(\D(\Eaa,\En),\D(\Ebb,\Em) \right)$ is an admissible
point. As shown in Proposition \ref{prop:alsojoint}, the joint
measurability of $\Eaa$ and $\Ebb$ implies that $\Ea$ and $\Eb$ are
jointly measurable.
Define $\va_0$ and $\vb_0$ to be the projections of the vectors
$\va$ and $\vb$, respectively, onto the plane spanned by $\vn$ and
$\vm$. Then
\begin{equation*}
\no{\va+\vb}\geq\no{\va_0+\vb_0},\qquad
\no{\va-\vb}\geq\no{\va_0-\vb_0},
\end{equation*}
and hence, $\E^{1,\va_0}$ and $\E^{1,\vb_0}$ are jointly measurable.
Using (\ref{eqn:distance-ab}) one finds that
\begin{equation}\begin{split}
\D(\E^{1,\va_0},\En) &\leq  \D(\Ea,\En) \leq \D_1,\\
\D(\E^{1,\vb_0},\Em) &\leq  \D(\Eb,\Em) \leq \D_2 .
\end{split}\end{equation}
We conclude that the best approximations are to be found from the
subset of covariant qubit observables, with vectors $\va$ and $\vb$
in the plane spanned by $\vn$ and $\vm$.\\

\noindent
(c)  If $(\D_1,\D_2)$ is an admissible point, then also $(\D'_1,\D'_2)$ is an admissible point whenever
$\D_i\leq\D'_i\leq\half$.\\
{\em Proof.}
In view of (b) it is sufficient to show the result for admissible points which have realizations
$\left(\D(\Ea,\En),\D(\Eb,\Em)\right)$. Thus let $\Ea$, $\Eb$ be two jointly measurable observables. Using Proposition \ref{prop:Paulsineq}, we note that also $\Ea$ and $\E^{1,r\vb}$ are jointly measurable for any $0\leq r\leq 1$. Since the function
\[
r\mapsto\D(\E^{1,r\vb},\Em)=\half\no{\vm-r\vb}
\]
is continuous, it takes all values between $\D(\Eb,\Em)$ and $\half$. We can similarly realize all values between $\D(\Ea,\En)$ and $\half$.
\qed\\

\noindent
(d) Observations (b) and (c) taken together entail
Proposition \ref{prop:admiss-realization}.\qed\\

\noindent
(e) The admissible region is a convex set.\\
{\em Proof.} Let $\left(\D_1,\D_2\right)$ and $\left(\D_1',\D_2'\right)$ be realized by
$(\alpha,\va), (\beta,\vb)$ and $(\alpha',\va'), (\beta',\vb')$ respectively. Let $t\in[0,1]$.
Then for
$(\alpha_t,\va_t):=(t\alpha+(1-t)\alpha',t\va+(1-t)\va')$ and
$(\beta_t,\vb_t):=(t\beta+(1-t)\beta',t\vb+(1-t)\vb')$, we obtain associated distances
$\D_{1,t}$ and $\D_{2,t}$ which satisfy
\[
\D_{k,t}\le t\D_k+(1-t)\D_k',\quad k=1,2.
\]
This together with (c) proves the claim.
\qed\\

\noindent
(f) The set of admissible points is closed.\\
{\em Proof.} The mapping
\begin{equation}
(\va,\vb)\mapsto (\D(\Ea,\En), \D(\Eb,\Em))=\half(\no{\va-\vn},\no{\vb-\vm})
\end{equation}
from $\R^3\times\R^3$ to $\R\times\R$ is continuous. The set of admissible points is the image of the compact set
\begin{equation}
\{ (\va,\vb)\in\R^3\times\R^3 \mid \no{\va}\leq 1, \no{\vb}\leq 1, \no{\va-\vb}+\no{\va+\vb}\leq 2 \},
\end{equation}
hence it is itself closed and  contains its boundary. This and (b) shows that for given
$\D_1\in[0,\half]$, there is a minimal number $\D_2^{\min}(\D_1)$ such that
all $\left(\D_1,\D_2\right)$ with $\D_2^{\min}(\D_1)\leq\D_2\leq\half$ are admissible pairs
while pairs with $\D_2<\D_2^{\min}(\D_1)$ are not admissible.\qed\\

\noindent
(g)  Since the admissible region is a convex set, the function $\D_1\mapsto\D_2^{\min}(\D_1)$ is convex and therefore continuous. Due to (a), the curve is symmetric under reflection with respect to the line $\D_1=\D_2$. We conclude that this function gives the lower boundary curve of the set of  admissible points, and that the points on this curve are admissible. Together with (e) and (f), this completes the proof of Proposition \ref{prop:convex}.

\vspace{10pt}

\noindent
{\bf Acknowledgement.} This work was initiated during T.H.'s visit at Perimeter Institute. Hospitality and support to both authors during this visit and to P.B. during the completion phase are gratefully acknowledged. T.H. acknowledges the support of the European Union project CONQUEST during the final phase of this work.

\end{document}